# Role of coil-crucible geometry in Czochralski bismuth germanate (BGO) crystal growth process: a thermal stress analysis

Hossein Khodamoradi[1,*], and Mohammad Hossein Tavakoli[1]

[1]Physics Department, Bu-Ali Sina University, Hamedan 65174, I.R. Iran

**Abstract**

A numerical model of 2D finite element in a steady-state level was developed for the electromagnetic field, heat distribution, and thermal stress expansion in an oxide Czochralski crystal growth system. The extended model was employed to compare the impact of different geometries of induction coils and crucibles in the growth process. Analysis of the results emphasizes the potential of the modified geometries in alteration of the electromagnetic field and heat distribution. Consequently, the optimization of crystal/melt interface shape, thermal stress accumulation, and distribution in the growing crystal can be achieved. Finally, the proposed approaches for thermal stress calculations were compared. The outcomes showed a qualitative agreement between two methods of thermoelastic stress analysis in the calculation of stress distribution in BGO crystal.

Tel: +82 10 2819 0660

E-mail: h.khodamoradi@basu.ac.ir, hossein.khodamoradi@gmail.com

# 1 Introduction

The bismuth germanate (BGO) crystals have been utilized in a wide range of applications from medical equipment (e.g. positron emission tomography) to high energy physics (e.g. low background spectrometry, high-energy calorimetry, and Compton suppression shield detectors). Suitable optical properties, reasonable crystal side smoothness, and preservable cylindrical shape are the factors that could make BGO crystals fit the mentioned applications. Conventional (high gradient) Czochralski technology with its key advantages can fulfill the conditions for growing such BGO crystals. Besides, this method comparably is less expensive and more flexible in terms of crystal size than Bridgman and Low Thermal Gradient Czochralski (LTG-Cz) methods [1]. Cz-BGO growth has also some issues like the lower melt consumption ratio, the higher tendency of crystal for faceting, and the growth in the helical form [2, 3, 4].

Induction-heating in the Cz method usually produces inhomogeneous heating in the metal crucible [5, 6, 7]. Such kind of heating in the growth system not only results in the convex melt-crystal interface but also may cause solidification of the melt in the vicinity of the crucible down part . Consequently, it lowers the coefficient of melt consumption [1, 8]. On the other hand, at the final stage of growth, touching the crucible bottom by the crystal tip would raise the risk of crucible deformation. There are two methods to prevent this problem: (1) using a bottom or lower heater [1, 6, 9], (2) changing the coil-crucible geometry [7, 10, 11]. Placing a bottom heater produces additional power in the crucible bottom which can increase the local temperature at the lower part of the melt. Such a phenomenon optimizes the coil-crucible combination and leads to produce more appropriate energy distribution in the crucible body as a passive solution.

Today, computer simulation is a practical tool to examine the operation of a crystal growth system under different conditions. Modeling of BGO crystal growth using the Czochralski method has a long term history both in the LG-Cz method (for example [4, 12, 13, 14, 15, 16]) or convectional Cz technique [17]. Computational study of an oxide crystal growth like BGO is complicated due to the high Prandtl number [18] of the melt and convoluted temperature pattern inside the melt [19]. Semi-transparent behavior combined with the other special optical properties of BGO makes these studies even more complicated [4, 13, 19].

Here we considered three different coil-crucible configurations and compared the obtained numerical results. Variation of the electromagnetic heat generation, temperature, and fluid velocity fields, crystal/melt interface, and thermal stress pattern in the crystal by modifying the heating system geometry are tracked during different stages of the growth. Afterward, the efficiency of the different coil-crucible configurations in reducing the crystallization front depth, decreasing the maximum values of developed thermal stress in the crystal, avoiding melt undercooling, and incrementing the efficiency of oxide Czochralski growth were addressed.

Three regarded configurations are: Case 1. Simple cylindrical both crucible and RF-coil, Case 2. Simple cylindrical crucible with L-shape coil, and Case 3. Crucible with round bottom corner and coil with a curved end (Fig. 1).

## 2 Mathematical formulation

### 2.1 Induction heating

The induction heating mathematical model was well defined in our previous paper which discussed the influence of crucible and coil geometry on induction heating [7].

### 2.2 Temperature and flow field

The fluid flow, continuity condition, and energy equations were assumed to be: (1) axially symmetric system and quasi-steady state, (2) laminar and incompressible Newtonian melt and gas flow while satisfying the Boussinesq approximation, (3) flat melt free surface, and (4) no viscous dissipation in the melt and gas used to calculate temperature and fluid flow field:

(a) Melt and gas flow

$$\rho \vec{V} \cdot \nabla \vec{V} = -\nabla p + \mu \nabla^2 \vec{V} + \rho \beta g (T - T_0) \hat{e}_z \tag{1}$$

(b) Continuity

$$\nabla \cdot \vec{V} = 0 \tag{2}$$

c) Energy

$$\alpha \nabla^2 T - \vec{V} \cdot \nabla T = 0 \text{ (heat transfer in the fluid parts)} \tag{3}$$

$$k \nabla^2 T + q_{\text{induction}} = 0 \text{ (heat transfer in the solid parts)} \tag{4}$$

$$k \nabla^2 T = \nabla \cdot \vec{q}_{\text{rad}} = \kappa \left( \int_{4\pi} I(\Omega) d\Omega - 4\pi I_b(T) \right) \text{ (radiation contribution inside the crystal)} \tag{5}$$

Where $\rho$ is the density, $\vec{V}$ is the vector of fluid velocity, $p$ is the pressure, $\mu$ is the dynamic viscosity, $\beta$ is the thermal expansion coefficient, $g$ is the acceleration of gravity, $T$ is the temperature, $\hat{e}_z$ is the unit vector along the pulling axes of Z, $\alpha$ is the thermal diffusivity, $k$ is the thermal conductivity, $q_{\text{induction}}$ is the heat source density, $\kappa$ is the absorption coefficient, and $q_{\text{rad}}$ is the radiation heat flux density. The first term inside the integral in Eq. (5) is heat flux in gray media where $I(\Omega)$ is the radiative intensity along the $\Omega$ direction and the second term is the black body radiative intensity multiplied by $4\pi$. In our approximation, radiative heat flux is added to the conductive heat flux in order to couple radiation in participating media (i.e. the grown crystal).

The boundary conditions are:

a) No-slip boundary condition at the fluid-solid boundaries

$$\vec{V} = 0 \tag{6}$$

b) Surface to surface radiation heat transfer at the gas-solid interfaces

$$-k_s \frac{\partial T_s}{\partial \hat{n}} = -k_g \frac{\partial T_g}{\partial \hat{n}} + \sigma \varepsilon_s \left(T_s^4 - T^4\right) \tag{7}$$

c) Surface to surface radiation heat transfer, Marangoni convection and no-slip boundary condition at the gas-melt interface

$$-k_m \frac{\partial T_m}{\partial \hat{n}} = -k_g \frac{\partial T_g}{\partial \hat{n}} + \sigma \varepsilon_m \left(T_m^4 - T^4\right) \tag{8}$$

$$\mu_m \frac{\partial u_m}{\partial \hat{n}} - \mu_g \frac{\partial u_g}{\partial \hat{n}} = \frac{\partial \gamma}{\partial \hat{\tau}} = \frac{\partial \gamma}{\partial T} \frac{\partial T}{\partial \hat{\tau}} \quad \text{(Marangoni effect)} \tag{9}$$

$$V_{\hat{n}} = 0 \tag{10}$$

d) Heat transfer considering the Latent heat generation proportional to the crystal growth rate, and crystal rotation as a boundary condition at the melt-crystal interface

$$k_m \frac{\partial T_m}{\partial \hat{n}} - k_c \frac{\partial T_c}{\partial \hat{n}} = -\rho_c H_f V_g \tag{11}$$

$$V_\varphi = r\omega_c \tag{12}$$

where the subscripts $s$, $g$, $m$, and $c$ mean solid, gas, melt, and crystal, respectively.

e) Gray wall radiation heat transfer at the crystal side surfaces [13, 20]

$$-k_c \frac{\partial T_c}{\partial \hat{n}} = \varepsilon_c W(T)\left(\sigma T_c^4 - T^{inc}\right) \tag{13}$$

$$V_\varphi = r\omega_c \tag{14}$$

f) Radiative and convective heat transfer at the outer surfaces of the chamber:

$$-k_{ins} \frac{\partial T_{ins}}{\partial \hat{n}} = h\left(T_{ins} - T_{amb}\right) + \sigma \varepsilon_{ins} \left(T_{ins}^4 - T_{amb}^4\right) \tag{15}$$

where $T_{amb}$ is the ambient temperature.

## 2.3 Thermal stress

The stress level in the crystals can effectively be determined by the temperature field during the growth process which consequently defines the content of crystal dislocations [19]. Analysis of the thermal stress for Cz growth of oxide crystals revealed the significant raise of thermal stress in semiconductor crystals like lithium niobite. It potentially can lead to producing crack during the crystal growth and cooling down process [21, 22, 23].

Thermoelastic stress fields inside as-grown crystals were calculated in terms of Von Mises stress. The detailed model explained and reported elsewhere [4]. Besides, it was shown that the dislocation density also can be correlated with the second derivative of the temperature profile [24]. Therefore, a formula for thermal stress introduced by Indenbom is hired to estimate the distribution of dislocations inside the crystals. This method is expressed by the following equation:

$$\sigma = \alpha_T E L^2 \left(\frac{\partial^2 T}{\partial Z^2}\right) \tag{16}$$

where $\alpha_T$ is the thermal expansion coefficient, $E$ is Young's modulus, $L$ is the crystal diameter, and $\sigma$ is thermal stress.

## 2.4 The calculation conditions

Three different kinds of coil-crucible geometry are considered in the work presented here which are;

Case 1. Simple cylindrical both crucible and RF-coil,
Case 2. Simple cylindrical crucible with L-shape coil, and

Case 3. Crucible with round bottom corner and coil with a curved end (Fig. 1).

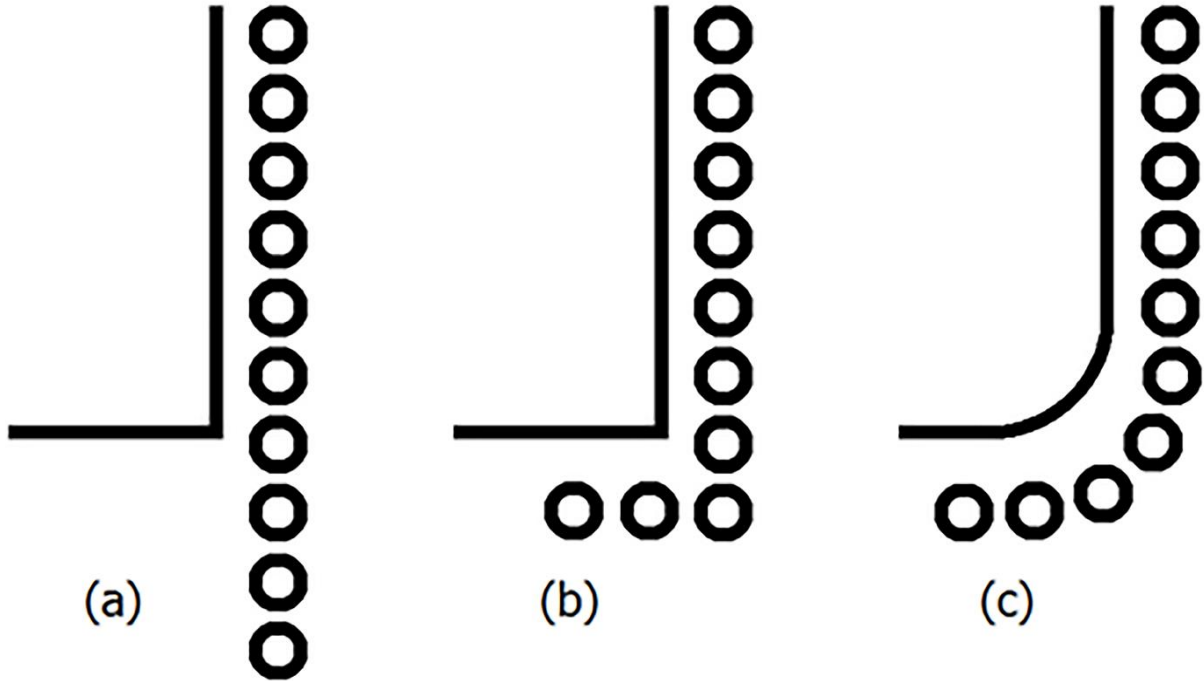

Fig. 1. Three considered coil-crucible configurations, (a) Case 1. Simple cylindrical both crucible and RF-coil, (b) Case 2. Simple cylindrical crucible with L-shape coil, and (c) Case 3. Crucible with round bottom corner and coil with a curved end.

The crystal pulling rate was set to 1 mm/H and the crystal rotation rate was fixed at 20 rev/min. Table 1 shows the operating parameter of the heating systems. The values of thermophysical properties used for calculation are presented in [4].

Table 1. Operating parameters of coil-crucible cases.

| Description (units) | Value |
|---|---|
| Crucible inner radius (mm) | 40 |
| Crucible wall thickness (mm) | 2 |
| Crucible inner height (mm) | 80 |
| Radius of the round bottom corner of crucible (mm) | 12 |
| Coil inner radius (mm) | 60 |
| Coil width (mm) | 2 |
| Distance between coil turns (mm) | 5 |
| Coil turns corner radius (mm) | 50 |

The BGO melt was taken to be completely opaque and BGO crystal was treated as a semi-transparent medium. Moreover, the melt-crystal interface was also assumed to be black. It should be noted that the gas flow was supposed to be laminar in most growth systems except in the high-pressure LEC of $A_3B_5$ crystals [25].

**2.5 Numerical method**

In this work, a 2D steady-state finite element method using the Newton-Raphson algorithm and direct solver MUMPS (Multifrontal Massively Parallel sparse direct Solver [26]) is employed to solve the governing equations for the whole calculations. The irregular shapes of the system (including crystal-melt interface and curved base of the crucible in Case 3) were handled using unstructured triangular body fitted elements. It was required to use carefully graded element distribution in order to achieve numerical stability and an adequately accurate approximation of the solutions in the calculation domains. To guarantee that the solution reveals the actual physical behavior, the element numbers have to be increased close to the interfaces and the areas where deep gradients of variables happen. The obtained numerical results were tested and verified by the simulation results of Refs. [27, 28]. Furthermore, the computational code and material properties were validated with respect to the experimental data in Ref. 4.

## 3 Results and discussion

Volumetric heat generation is shown in Fig. 2 for crucible domains of the mentioned cases [7]. Although the distribution of heat generation in the crucible does not change during the growth process, its intensity varies proportionally to the melt height and requires thermal condition (Table 2). As a result, the intensity and the total volumetric heat decreases by increasing the crystal length in every case. This reasonable reduction is caused by lowering the melt volume and so reaching the smaller power to the melting point.

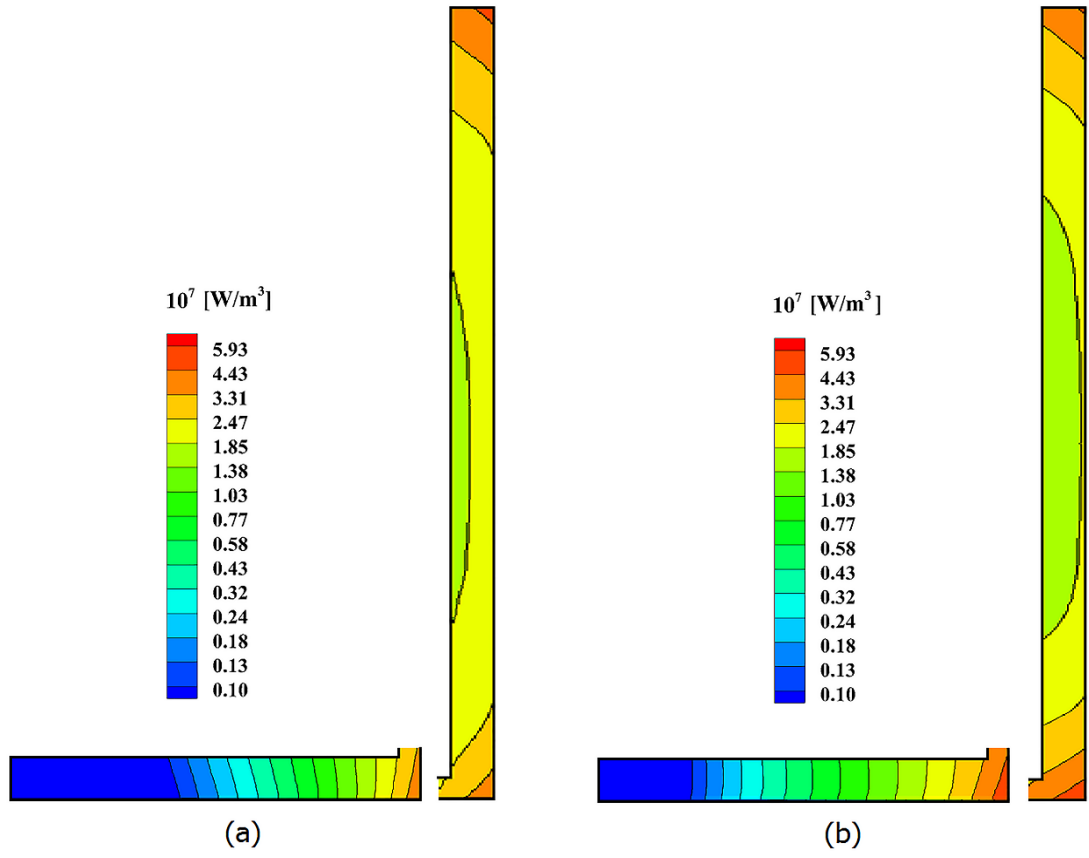
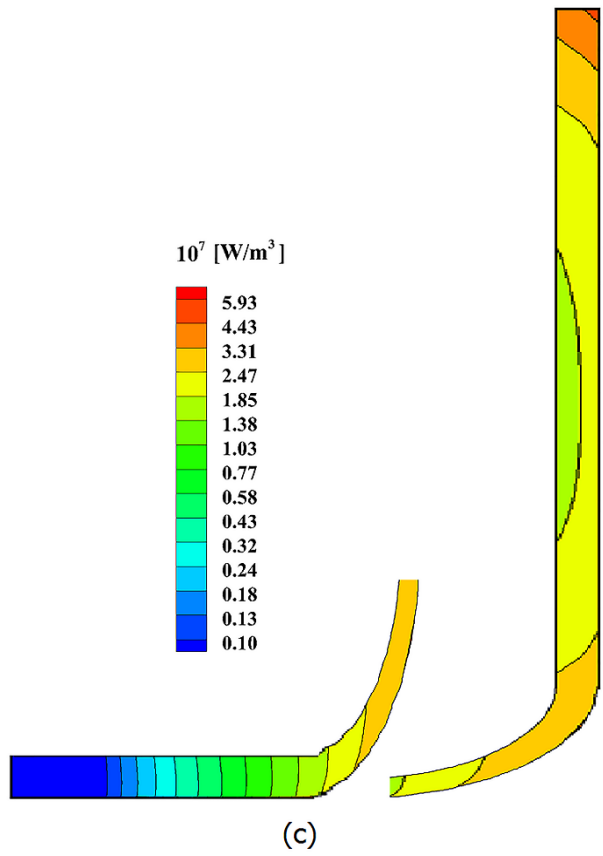

Fig. 2. Crucibles heat generation patterns (a) Case 1, (b) Case 2, and (c) Case 3 (the crucible wall and bottom parts are separately magnified for a better demonstration).

Table 2. Total volumetric heat in the crucible for all cases.

| Description (units) | Value |
|---|---|
| Crucible inner radius (mm) | 40 |
| Crucible wall thickness (mm) | 2 |
| Crucible inner height (mm) | 80 |
| Radius of the round bottom corner of crucible (mm) | 12 |
| Coil inner radius (mm) | 60 |
| Coil width (mm) | 2 |
| Distance between coil turns (mm) | 5 |
| Coil turns corner radius (mm) | 50 |

Fig. 3 shows the temperature and flow fields of the whole Cz puller for the three considered cases. There is a single melt vortex that forms from melt raising along the crucible wall and two gas vortices (a small and a large) with opposite rotation in all cases. Gas flow just above the melt is identified by a small and strong eddy that has the same direction of melt surface flow. The heat that is transferred from the crucible wall to the grown crystal affects the temperature gradient.

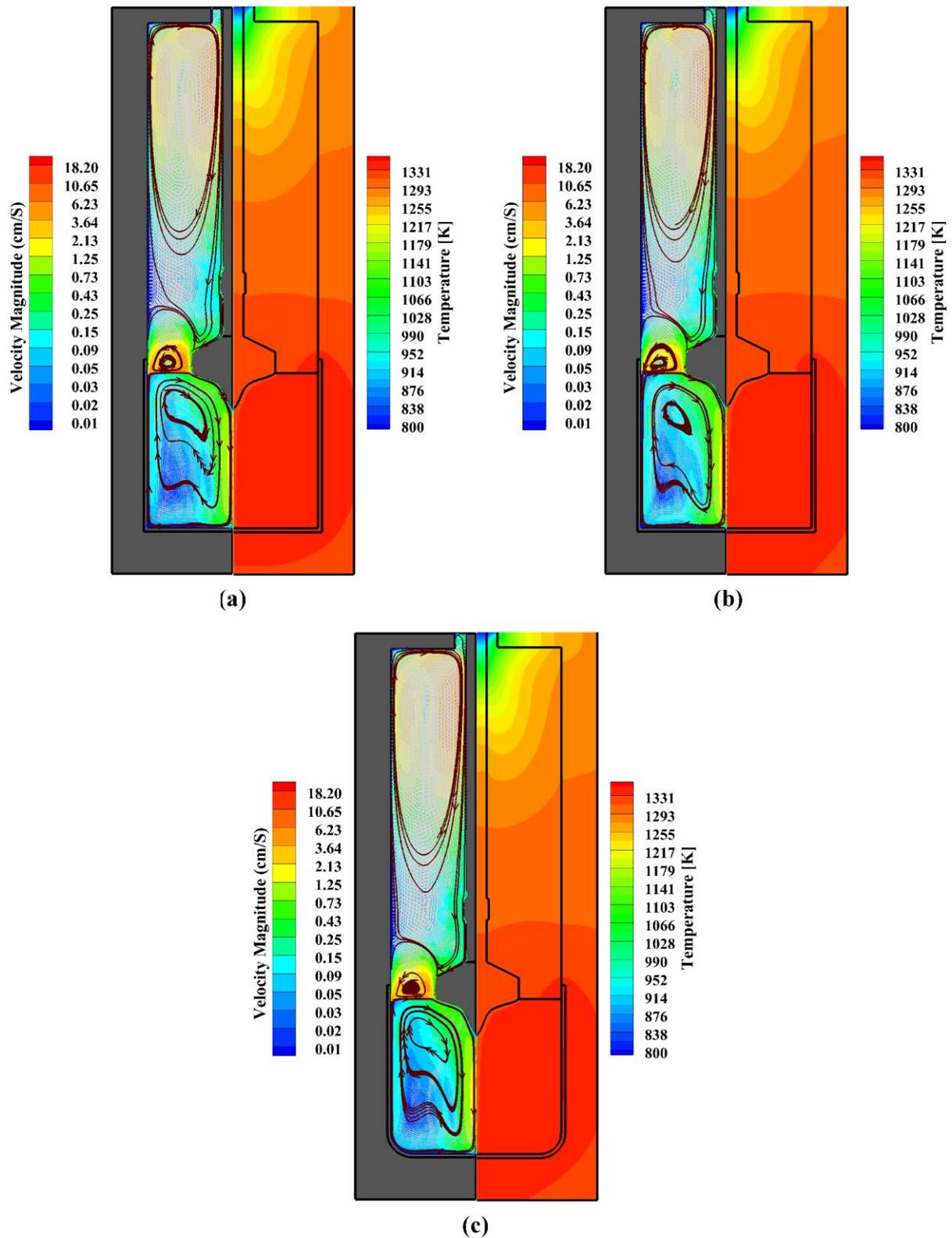

Fig. 3. Distribution of temperature (right-hand side) and velocity (left-hand side) in the whole growth setup (a) Case 1, (b) Case 2 and (c) Case 3.

Melt convection, characterized as a large single eddy, flows along with the crystal/melt interface and toward the crucible bottom. It replicates the intrinsic strong natural convection in oxide crystal melts like sapphire (Fig. 4) [29]. This flow behavior of the melt has a significant influence on the melt-crystal interface deflection, and the temperature distribution and transportation in the molten BGO. Even by close attention to the crystal/melt interface, there is not any opposite flow vorticity direction or significant turbulence in the melt flow. Therefore, the coupling natural and Marangoni convections dominate the melt convection and suppress the influence of force convection caused by the crystal rotation.

On the other hand, the strong natural convection and the weak conductive heat transfer of BGO melt formed the convoluted temperature patterns of the melt. The maximum temperature, maximum fluid velocity, and center of the main vortex in Case 2 place notably in lower position than the other two cases which could be examined to approximate the strength of buoyancy convection. It is also worth mentioning that the maximum value of the fluid velocity inside the melt in Case 2 is much lower than the other cases (0.7 cm/s versus 1.45 cm/s and 1.51 cm/s respectively in Case 1 and Case 3). It is considered as the second main parameter to lead comparably weaker convection in Case 2.

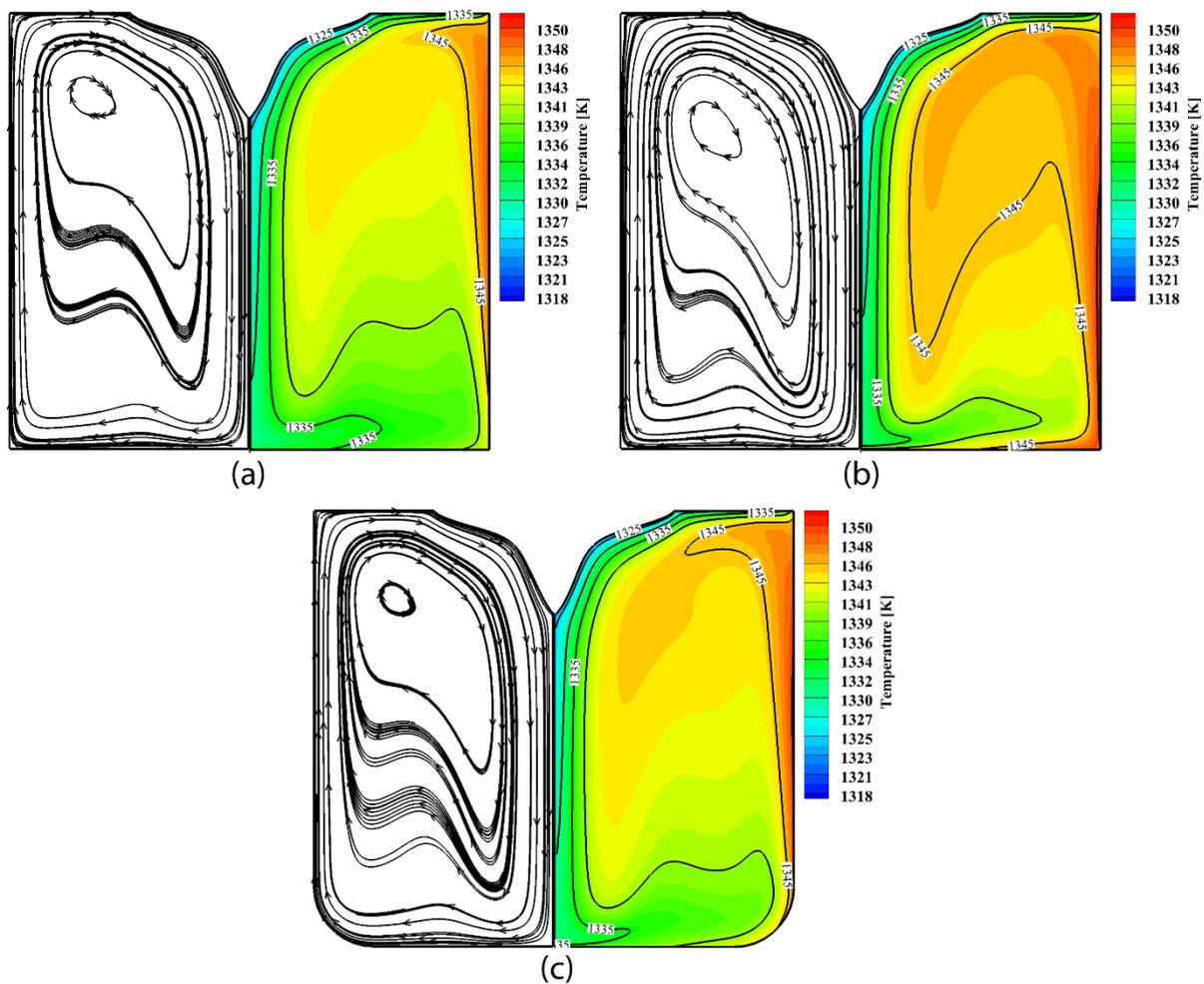

Fig. 4. Distribution of temperature (right-hand side) and streamline (left-hand side) (a) Case 1, (b) Case 2 and (c) Case 3.

Generally, the crystal/melt interface tends to be convex and deflected toward the melt in the growth process of BGO crystals due to the lower thermal conductivity of BGO melt comparing to its crystal [19]. Two reasons were mentioned in different literature for the deep-convex shape of growth front in BGO crystals: (1) domination of natural convection, and (2) transfer of bulk radiative heat from the semi-transparent crystals [4, 30]Comparing the height of crystal deflection toward the melt in Table 3 reveals that the lowest deflection occurred in Case 2. The crystal fronts in Case 1 and Case 3 show nearly similar deflection height and shape of the convexity. The similar calculations of crystal front for larger crystals of 5c m and 15 cm in lengths manifest that the melt/crystal interfaces move deeper toward the melt by about 20% and 60% respectively compared to the crystals of 1 cm in deflection height (Table 3). The later results can be explained as the consequence of lowering the melt level and growing the crystal which correspondingly cause less strong convection in the melt. In every considered stage of the crystal growth in different lengths, Case 2 has the lowest interface deflection height. Also, the difference of maximum and minimum values of the deflection height between the considered cases at the same stage changes from ~8% for 1 cm crystals to ~24% for 15 cm crystal which potentially enhances the risk of the core phenomena, microcracking and facetted interface [19].

In our calculation for case 1 with 15cm crystallization stage, the parasite crystals were recognised at the down part of the melt which are a proof of the importance of selecting proper heating setup.

Table 3. Crystal/Melt deflection height for all cases.

| Crystal height (cm) | Case 1 (mm) | Case 2 (mm) | Case 3 (mm) |
|---|---|---|---|
| 1 | 17.6 | 21.2 | 28.6 |
| 5 | 16.4 | 18.9 | 23.0 |
| 15 | 17.7 | 21.7 | 27.2 |

Fig. 5 compares the distribution of Von Mises stress in the grown crystal considered as the first-order derivative of thermal stress and the Indenbom's approach of the second-order stress approximation for 1cm crystals in the studied cases. The values of both computed stress calculation methods are plotted show the significant differences in the magnitude of the predicted stress between the results of these two methods. Also, some similarities were found like both methods suggested the concentration of high stress at the crystal top, the crystal-melt-gas triple point (as hot spots) and the crystal/melt interface while the thermal stress was attenuated to its lowest value at the crystal core area.

In all three cases, the Von Mises stress analysis demonstrates that the absolute maximum value of the stress is located near the seed, whereas the second-order stress analysis reveals a local maximum at the seed adjacent and the absolute maximum at the triple point. Between all three cases shown in Fig. 5, Case 2 has a smoother and more homogeneous stress gradient than the crystal axisymmetric axis to the crystal sidewall. Although, the highest stress accumulation at the crystal/melt interface in both analysis methods is identified for Case 2.

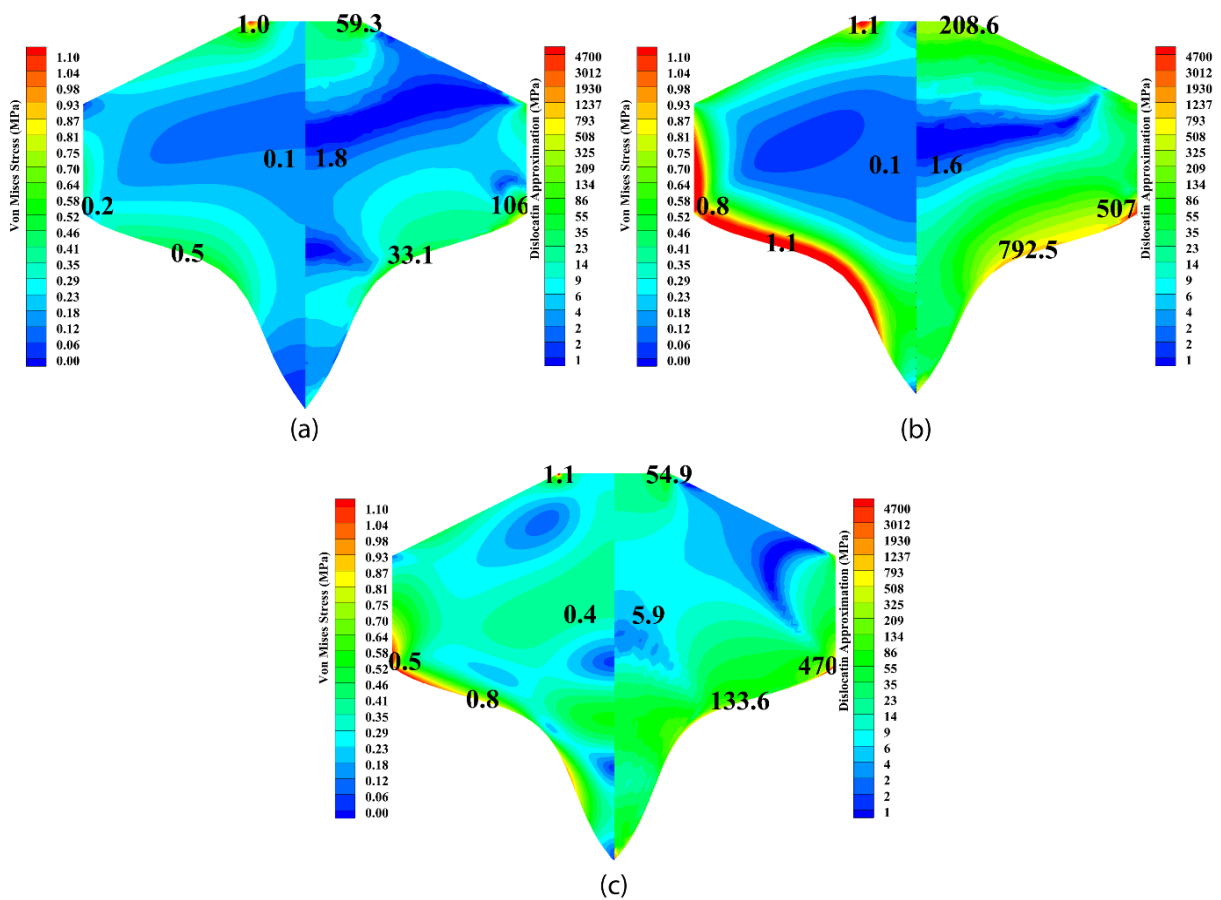

Fig. 5. Von Mises stress distribution as the first-order derivative of thermal stress (left-hand side of each graph) and Indenbom approach as the second-order stress approximation (left-hand side) for (a) Case 1, (b) Case 2 and (c) Case

In this study, the maximum value of Von Mises stress is significantly increasing as the crystal reaches the higher lengths, whereas this value remains in the same range between different considered cases at the same height. For 1cm crystal height, the maximum value of Von Mises stress is in the range of 1-1.2 MPa. For 5cm crystal, the maximum value boosts and peaks at 5.15 MPa in Case 3 while the lowest maximum value occurs in Case 1 at 4.21 MPa. Finally, for the 15cm BGO crystals, thermal stress is increased by almost six times and reaches the range of 29-30 MPa in the three cases.

The reported tensile strength for BGO crystals has a comparably low value of about 23.2 MPa [31]. Therefore, at the final stage of the growth process where the maximum value of Von Mises stress exceeds the tensile strength limit, crack defects can be developed in the studied system. Although, the area of accumulation of such a high stress is a small spot at the edge of seed connected to the crystal shoulder, producing any crack defects at this point would come with the risk of its propagation through the grown crystal crown and consequently degrading the crystal quality.

To the best of the author's knowledge, critical resolved shear stress (CRSS) value of the BGO crystals has not been reported in any wide accessible source. Therefore, reaching a pattern of dislocation density using the calculated Von Mises stress was not possible. In addition, because of the lack of data available to the authors, applying more reliable methods like alexander-haasen sumino analysis was also not part of our research.

## 4 Conclusion

A finite element simulation was employed to study the effect of coil/crucible configurations in the RF-heating setup of a Cz oxide crystal growth process. Two different approaches were investigated to model thermal stress patterns in the growing crystals. Revolusion of the crystal/melt interface and thermal stress enhancement were examined during different stages of the BGO process.

The simulation results show that the changing shape of the electromagnetic heating setup elements can reduce the crystal/melt deflection more than 24% in BGO growth. It can also prevent the formation of parasite crystals at the melted bottom and early reaching of the crystal/melt interface to the crucible bottom.

Comparison of first- and second-order thermal stress calculations reveals that both methods indicate a dramatic change of stress distribution in crystal while exhibiting high stress regions at the crystal/melt interface, down part of the crystal rim, and close to the seed. The pattern of thermal stress inside the crystal can differ from radially more homogeneous to heterogeneous by changing the heating setups. There is a remarkable enhancement of maximum thermal stress as the crystal grows in length. In addition, the thermal stress can exceed the tensile strength of BGO crystals when the crystal/melt interface is close to the crucible bottom.

In conclusion, the variation in coil/crucible geometry as a passive solution to optimize the heating conditions of semi-transparent oxide crystals with a highly convex melt/crystal interface can effectively reduce the crystallization front deflection. In order to achieve a flat interface, investigating the influence of a magnetic flux concentrator (MFC) along a proper choice of heating setup geometry is highly suggested. Furthermore, reducing the maximum thermal stress accumulation inside the crystal during the growth process and applying an active after-heater, in order to decrease the heat loss from the crystal body by internal radiation, can be very promising. Last but not least, a 3D model of thermal stress and an Alexander Haasen calculation were performed to investigate the generation and propagation of dislocations. These can help better understanding of the dislocation dynamics in BGO crystals which needs more information from the growers and producers.

## References

[1] A. V. Kolesnikov, E. P. Galenin, O. T. Sidletskiy and V. V. Kalaev, "Optimization of heating conditions during Cz BGO crystal growth," *Journal of Crystal Growth,* vol. 407, p. 42–47, 2014.

[2] B. T. O. K. V. B. B. K. a. E. P. S.F. Burachas, *Functional Materials,* vol. 4, p. 305, 1997.

[3] R. Voszka, G. Gévay, I. Földvári and S. Keszthelyi-Lándori, "Growth and characterization of Bi4Ge3O12 single crystals," *Acta Physica Academiae Scientiarum Hungaricae,* vol. 53, p. 7–13, 1982.

[4] S. Omid and M. H. Tavakoli, "Effect of the ceramic tube shape on global heat transfer, thermal stress and crystallization front in low thermal gradient (LTG) Czochralski growth of scintillating BGO crystal," *Materials Research Express,* vol. 5, p. 105507, 2018.

[5] M. H. Tavakoli, A. Ojaghi, E. Mohammadi-Manesh and M. Mansour, "Influence of coil geometry on the induction heating process in crystal growth systems," *Journal of Crystal Growth,* vol. 311, p. 1594–1599, 2009.

[6] M. H. Tavakoli, E. Mohammadi-Manesh and A. Ojaghi, "Influence of crucible geometry and position on the induction heating process in crystal growth systems," *Journal of Crystal Growth,* vol. 311, p. 4281–4288, 2009.

[7] H. Khodamoradi, M. H. Tavakoli and K. Mohammadi, "Influence of crucible and coil geometry on the induction heating process in Czochralski crystal growth system," *Journal of Crystal Growth,* vol. 421, p. 66–74, 2015.

[8] M. Honarmandnia, M. H. Tavakoli and H. Sadeghi, "Global simulation of an RF Czochralski furnace during different stages of germanium single crystal growth," *CrystEngComm,* vol. 18, p. 3942–3948, 2016.

[9] K. Mazaev, V. Kalaev, E. Galenin, S. Tkachenko and O. Sidletskiy, "Heat transfer and convection in Czochralski growth of large BGO Crystals," *Journal of Crystal Growth,* vol. 311, p. 3933–3937, 2009.

[10] K. Takagi and T. Fukazawa, "Effect of growth conditions on the shape of Bi4Ge3O12 single crystals and on melt flow patterns," *Journal of Crystal Growth,* vol. 76, p. 328–338, 1986.

[11] M. Honarmandnia, M. H. Tavakoli and H. Sadeghi, "Global simulation of an RF Czochralski furnace during different stages of germanium single crystal growth, part II: to investigate the effect of the crucible's relative position against the RF coil on the isotherms, flow fields and thermo-elastic stresses," *CrystEngComm,* vol. 19, p. 576–583, 2017.

[12] I. Y. Evstratov, S. Rukolaine, V. S. Yuferev, M. G. Vasiliev, A. B. Fogelson, V. M. Mamedov, V. N. Shlegel, Y. V. Vasiliev and Y. N. Makarov, "Global analysis of heat transfer in growing BGO crystals (Bi4Ge3O12) by low-gradient Czochralski method," *Journal of Crystal Growth,* vol. 235, p. 371–376, 2002.

[13] V. S. Yuferev, O. N. Budenkova, M. G. Vasiliev, S. A. Rukolaine, V. N. Shlegel, Y. V. Vasiliev and A. I. Zhmakin, "Variations of solid-liquid interface in the BGO low thermal gradients Cz growth for diffuse and specular crystal side surface," *Journal of Crystal Growth,* vol. 253, p. 383–397, 2003.

[14] V. D. Golyshev and M. A. Gonik, "Heat transfer in growing Bi4Ge3O12 crystals under weak convection: II - Radiative-conductive heat transfer," *Journal of Crystal Growth,* vol. 262, p. 212–224, 2004.

[15] O. N. Budenkova, M. G. Vasiliev, V. N. Shlegel, N. V. Ivannikova, R. I. Bragin and V. V. Kalaev, "Comparative analysis of the heat transfer processes during growth of Bi12GeO20 and Bi4Ge3O12 crystals by the low-thermal-gradient Czochralski technique," *Crystallography Reports,* vol. 50, p. 100–105, 2005.

[16] M. G. Vasiliev, V. M. Mamedov, S. A. Rukolaine and V. S. Yuferev, "Heat source optimization in a multisection heater for the growth of bismuth germanate crystals by the low-gradient Czochralski method," *Bulletin of the Russian Academy of Sciences: Physics,* vol. 73, p. 1406–1409, 2009.

[17] V. M. Mamedov and V. S. Yuferev, "Time-dependent model of the growth of oxide crystals from melt by the Czochralski method," *Bulletin of the Russian Academy of Sciences: Physics,* vol. 73, p. 1402–1405, 2009.

[18] Y. R. Li, L. Zhang, L. Zhang and J. J. Yu, "Experimental study on Prandtl number dependence of thermocapillary-buoyancy convection in Czochralski configuration with different depths," *International Journal of Thermal Sciences,* vol. 130, p. 168–182, 2018.

[19] B. K. P. V. a. D. M. Dhanaraj G, Springer Handbook of Crystal Growth, Berlin: Springer, 2010.

[20] Z. Guo, S. Maruyama and T. Tsukada, "Radiative heat transfer in curved specular surfaces in czochralski crystal growth furnace," *Numerical Heat Transfer; Part A: Applications,* vol. 32, p. 595–611, 1997.

[21] T. Tsukada, K. Kakinoki, M. Hozawa, N. Imaishi, K. Shimamura and T. Fukuda, "Numerical and experimental studies on crack formation in LiNbO3 single crystal," *Journal of Crystal Growth,* vol. 180, p. 543–550, 1997.

[22] N. Miyazaki, H. Uchida, T. Tsukada and T. Fukuda, "Quantitative assessment for cracking in oxide bulk single crystals during Czochralski growth: Development of a computer program for thermal stress analysis," *Journal of Crystal Growth,* vol. 162, p. 83–88, 1996.

[23] M. Kobayashi, T. Tsukada and M. Hozawa, "Effect of internal radiation on thermal stress fields in CZ oxide crystals," *Journal of Crystal Growth,* vol. 241, p. 241–248, 2002.

[24] V. L. Indenbom, "Ein Beitrag zur Entstehung von Spannungen und Versetzungen beim Kristallwachstum," *Kristall und Technik,* vol. 14, p. 493–507, 1979.

[25] D. K. Ofengeim and A. I. Zhmakin, "Industrial challenges for numerical simulation of crystal growth," *Lecture Notes in Computer Science (including subseries Lecture Notes in Artificial Intelligence and Lecture Notes in Bioinformatics),* vol. 2657, p. 3–12, 2003.

[26] [Online]. Available: http://mumps.enseeiht.fr/.

[27] N. Kobayashi, "Hydrodynamics in Czochralski growth-computer analysis and experiments," *Journal of Crystal Growth,* vol. 52, p. 425–434, 1981.

[28] M. H. Tavakoli, S. Omid and E. Mohammadi-Manesh, "Influence of active afterheater on the fluid dynamics and heat transfer during Czochralski growth of oxide single crystals," *CrystEngComm,* vol. 13, p. 5088–5093, 2011.

[29] C. Stelian, G. Sen and T. Duffar, "Comparison of thermal stress computations in Czochralski and Kyropoulos growth of sapphire crystals," *Journal of Crystal Growth,* vol. 499, p. 77–84, 2018.

[30] J. Banerjee and K. Muralidhar, "Role of internal radiation during Czochralski growth of YAG and Nd:YAG crystals," *International Journal of Thermal Sciences,* vol. 45, p. 151–167, 2006.

[31] K. Nakajima, "Mechanical design of Hard X-ray Imager and Soft," 2011.